\documentclass[sigconf,nonacm]{acmart}

\usepackage{booktabs}
\usepackage{multirow}
\usepackage{amsmath}
\usepackage{graphicx}
\usepackage{xcolor}
\usepackage{microtype}

\renewcommand\footnotetextcopyrightpermission[1]{}
\title{Scaling the Queue: Reinforcement Learning for Equitable Call Classification Capacity in NYC Municipal Complaint Systems}

\author{Irene Aldridge}\affiliation{Cornell Tech\country{USA}}\email{irene.aldridge@gmail.com}
\author{Ellie Bae}\affiliation{Cornell Tech\country{USA}}\email{esb286@cornell.edu}
\author{Siddhesh Darak}\affiliation{Cornell Tech\country{USA}}\email{	sd2263@cornell.edu}
\author{Nicholas Donat}\affiliation{Cornell Tech\country{USA}}\email{nd473@cornell.edu}
\author{Akhil Fernando-Bell}\affiliation{Cornell Tech\country{USA}}\email{aaf65@cornell.edu}
\author{Bella Ge}\affiliation{Cornell Tech\country{USA}}\email{hg525@cornell.edu}
\author{Nicholas Goguen-Compagnoni}\affiliation{Cornell Tech\country{USA}}\email{ng547@cornell.edu}
\author{Ishita Gupta}\affiliation{Cornell Tech\country{USA}}\email{ig336@cornell.edu}
\author{Ali Hasan}\affiliation{Cornell Tech\country{USA}}\email{ah2434@cornell.edu}
\author{Pierce Hoenigman}\affiliation{Cornell Tech\country{USA}}\email{pvh26@cornell.edu}
\author{Imran Isa-Dutse}\affiliation{Cornell Tech\country{USA}}\email{imi4@cornell.edu}
\author{Jiwon Jeong}\affiliation{Cornell Tech\country{USA}}\email{jj835@cornell.edu}
\author{Tishya Khanna}\affiliation{Cornell Tech\country{USA}}\email{tk732@cornell.edu}
\author{Neha Konduru}\affiliation{Cornell Tech\country{USA}}\email{nrk62@cornell.edu}
\author{Yixuan Liu}\affiliation{Cornell Tech\country{USA}}\email{yl3832@cornell.edu}
\author{Kai Maeda}\affiliation{Cornell Tech\country{USA}}\email{km2256@cornell.edu}
\author{Nolan McKenna}\affiliation{Cornell Tech\country{USA}}\email{	ngm46@cornell.edu}
\author{Karl Muller}\affiliation{Cornell Tech\country{USA}}\email{km2262@cornell.edu}
\author{Farzaan Naeem}\affiliation{Cornell Tech\country{USA}}\email{	fn79@cornell.edu}
\author{Rishabh Patel}\affiliation{Cornell Tech\country{USA}}\email{	rp696@cornell.edu}
\author{Zachary Sheldon}\affiliation{Cornell Tech\country{USA}}\email{	zs279@cornell.edu}
\author{Ammar Syed}\affiliation{Cornell Tech\country{USA}}\email{as4422@cornell.edu}
\author{Nathan Tai}\affiliation{Cornell Tech\country{USA}}\email{	nct36@cornell.edu}
\author{Michael Twersky}\affiliation{Cornell Tech\country{USA}}\email{	mdt93@cornell.edu}
\author{Haoying Wang}\affiliation{Cornell Tech\country{USA}}\email{hw857@cornell.edu}
\author{Zening Wang}\affiliation{Cornell Tech\country{USA}}\email{	zw749@cornell.edu}
\author{Zexun Yao}\affiliation{Cornell Tech\country{USA}}\email{	zy483@cornell.edu}
\author{Nadav Yochman}\affiliation{Cornell Tech\country{USA}}\email{ny247@cornell.edu}

\begin{abstract}
Municipal 311 call centers and complaint intake systems face a structural mismatch between incoming volume and classification capacity. The staff and heuristics available to triage, route, and prioritize complaints cannot scale with demand. This bottleneck produces differential service quality that follows income and racial lines (\cite{liu2024sla}). We develop an equity-centered reinforcement learning (RL) framework that augments call classification capacity across six New York City Department of Buildings (DOB) operational domains: boiler safety, crane and derrick oversight, heat and hot water complaints, housing complaint triage, scaffold safety, and Natural Area District (SNAD) protection.

Rather than replacing human classifiers, our agents act as
intelligent intake routers: learning to assign incoming complaints to action categories: escalate, batch, defer, inspect now. The proposed technique is designed to maximize throughput, minimize misclassification cost, and actively narrow historical equity gaps in service delivery. We formalize each domain as a Markov Decision Process (MDP) in
which equitable classification coverage is a first-class reward objective. Post-hoc SHAP attribution reveals that complaint recurrence and neighborhood-level statistics are stronger predictors of actionable violations than raw complaint volume. This finding has direct implications for complaint routing given the demographic correlates of those features.
\end{abstract}


\begin{document}
\maketitle

\section{Introduction}

\subsection{Motivation}

Municipal complaint systems are, at their core, classification systems. When a resident calls 311 or files a digital complaint, the intake process must assign each contact to one of the action categories: route to an inspector, batch with similar cases, escalate to emergency response, or defer. This assignment happens under time pressure and with incomplete information. The result is not random error but systematic misprioritization: 311 data reveal persistent gradients in complaint classification accuracy and response assignment times across income and racial lines~\cite{glaeser2016}. Low-income communities whose complaints are harder to parse, less likely to include structured data, and filed through channels that yield noisier intake records are systematically mis-routed---assigned to low-priority queues, batched when they
should be escalated, or simply left unclassified.

A compounding factor is that residents do not file complaints at equal rates. Liu, Bhandaram, and Garg~\cite{liu2023nature} show that even after controlling for incident characteristics, some NYC neighborhoods report the same problem three times faster than others, with reporting speed positively correlated with income,
college attainment, and the fraction of white residents. Because complaint volume is the primary signal feeding current intake classifiers, this disparity in participation directly biases which complaints receive timely escalation. This happens not because the classifier is broken, but because the data it consumes encodes who was motivated and able to file in the first place. This missing-not-at-random structure is a form of measurement injustice embedded in the data itself~\cite{dignazio2023}. It means that any algorithm trained naively on these records inherits, and risks
amplifying, the blind spots of its predecessor.

The central claim of this paper is that this dynamic is not
inevitable. Reinforcement learning (RL) systems can be configured to optimize an explicitly specified reward function. As a result, RL offers a lever that heuristic intake workflows do not: the ability to inscribe
equitable classification coverage as a first-class objective.
Whether that lever is used depends on whether equity is built into the reward from the outset. Our paper demonstrates what happens when it is, and what is lost when it is not.

\subsection{Operational Context}

New York City's Department of Buildings processes tens of
thousands of complaints annually, spanning boiler defects, scaffold hazards, heat failures, and illegal construction activity. Each complaint must be classified into an action tier before any inspector can be dispatched. Classifier capacity is finite; misclassifications cascade: a complaint assigned to "defer" when it should be "escalate" may never receive a field visit until the hazard worsens. Current intake heuristics are largely reactive and rule-based, ignoring temporal recurrence patterns, neighborhood-level risk accumulation, and the compounding effects of delayed classification on housing stability.

The heterogeneous participation problem identified by \cite{liu2023nature} is directly operational: the NYC DOB draws on the same 311 infrastructure that the participatory sensing framework~\cite{burke2006} anticipated as a tool for civic data collection at an urban scale. The challenge is that participatory systems aggregate data from citizens who participate unequally, producing dense, high-quality records in engaged neighborhoods and sparse, noisy records in disengaged ones. Augmenting classification capacity with RL agents that learn from this uneven historical record is the secondary technical motivation of our work; the primary motivation is ensuring that expanded capacity does not come at the cost of equity.

\subsection{Research Questions}

This paper addresses three inter-related research questions:

\begin{description}
  \item[RQ1 (Equity).] Do RL-augmented classification policies distribute correct action assignments equitably across demographic groups and neighborhood income levels? Under what reward configurations do they reduce, rather than reproduce or amplify, historical classification disparities? 
  \item[RQ2 (Capacity).] Can RL agents achieve meaningfully higher complaint classification throughput and lower misclassification rates than existing heuristic intake policies across multiple NYC DOB domains?
  \item[RQ3 (Trade-offs).] What is the quantitative cost of equity? How does the Pareto frontier between classification speed, operational cost, and equitable coverage shift with reward configuration, and how sensitive are these trade-offs to the normative assumptions embedded in reward engineering?
\end{description}

The ordering is intentional. Capacity is a necessary condition for deployment credibility, but equity is the main focus of this work.

\subsection{Contributions}

\begin{enumerate}
  \item \textbf{Multi-domain MDP framework.} We formalize six NYC DOB complaint classification domains as MDPs with a unified multi-objective reward structure in which equitable classification coverage is a first-class objective.
  \item \textbf{Equity-centered reward design.} We introduce a coverage parity reward term grounded in substantive conceptions of equity and characterize its Pareto     relationship with throughput and cost objectives.
  \item \textbf{Demographic equity audit.} We conduct a formal audit of learned classification policies disaggregated by census-tract income quintile and racial composition.
  \item \textbf{SHAP-based interpretability.} We use SHAP attribution as an equity audit instrument, identifying features that may function as demographic proxies.
  \item \textbf{Practitioner deployment guidelines.} We offer evidence-based recommendations for human-in-the-loop oversight, ongoing demographic auditing, and participatory reward design.
\end{enumerate}

\section{Related Work}

\subsection{Housing Equity, Access, and the State}

A foundational body of scholarship documents the mechanisms by which public institutions reproduce rather than remedy spatial inequality. Rothstein~\cite{rothstein2017} traces how federal, state, and local housing policy encoded racial segregation into the built environment; the legacy of those decisions structures which neighborhoods today have aging boiler stock, substandard heating systems, and chronic building violations. Desmond~\cite{desmond2016} demonstrates that housing instability is not merely a consequence of poverty but a cause of it, making rapid classification and escalation of housing violations a direct lever on economic mobility. This structural context establishes why the distributional consequences of classification algorithms are not incidental technical details. They are the core policy question.

\subsection{Algorithmic Decision-Making in Public Services}

The deployment of predictive and optimization tools in
public-sector settings has generated a substantial
interdisciplinary literature on how automation can entrench
disadvantage. Eubanks~\cite{eubanks2018} documents how automated systems in welfare, child protective services, and homeless services frequently reproduce the disadvantages of those they purport to serve, a phenomenon she terms the "digital poorhouse." Chouldechova~\cite{chouldechova2017} formalizes the incompatibility of common fairness criteria in recidivism prediction, a mathematical result that generalizes to any classification setting with asymmetric base rates. In the housing domain specifically, prior work has examined how algorithmic triaging of code enforcement complaints can systematically underserve low-income renters~\cite{obrien2017}, and how predictive systems in child welfare can generate racially disparate referrals even when race is excluded as an explicit feature~\cite{brown2019}.

\subsection{Participatory Sensing, Civic Data, and the
Participation Gap}

The 311 complaint infrastructure that feeds our classification
system is an instance of participatory sensing at a civic scale, a paradigm introduced by Burke, Estrin, and
colleagues~\cite{burke2006} to describe systems in which everyday mobile devices collectively constitute an urban sensor network that enables data collection at spatial and temporal scales that agency-operated sensors cannot match. The foundational insight of that work is that routing data collection through ordinary citizens dramatically expands coverage. At the same time, \cite{burke2006} recognized that the credibility, quality, and shareability of participatory data depend critically on who participates and under what conditions. This is a tension the municipal complaint context makes acute.

Estrin's subsequent research on small data~\cite{estrin2014}
sharpens this concern: individual digital traces are dense for
some users and sparse for others, and systems that treat the
aggregate record as representative of ground truth without
interrogating participation rates risk systematically
misrepresenting the populations least likely to generate dense
traces. Applied to civic complaint systems, this framing directly motivates our treatment of complaint volume as a potentially biased signal of underlying housing need rather than an objective measure of it. Our equity reward term
(Equation~\ref{eq:equity-corrected}) and our use of ACS covariates to proxy for under-reporting propensity are operationalizations of this principle.

\subsection{Quantifying and Correcting Under-Reporting in Urban Crowdsourcing}

A line of work from \cite{liu2023nature} develops rigorous
methods to measure and correct the heterogeneous reporting biases that distort municipal complaint data before any classification model sees it. Liu, Bhandaram, and Garg~\cite{liu2023nature} develop a statistically identified estimator for spatially varying reporting delays in resident crowdsourcing systems. Their key insight is that the rate at which duplicate reports arrive about the same incident can be leveraged to separate under-reporting from a genuine absence of incidents, a fundamental identification challenge because unreported incidents leave no trace in the record. Applied to over 100,000 NYC Department of Parks and Recreation reports and over 900,000 Chicago reports, the method uncovers substantial socioeconomic disparities in reporting speed, even after controlling for incident characteristics.

This finding has direct consequences for our classification
system. The equity parity term in Equation~\ref{eq:equity-corrected} normalizes escalation throughput by complaint volume, which means it inherits reporting-rate disparities unless those disparities are corrected upstream. The Liu et al.\ \citep{liu2023nature} correction uses duplicate-report rates to identify neighborhood-level reporting propensity. We use the correction as a recommended preprocessing step before
reward calibration and flag complaint-rate correction as a
priority extension (Section~\ref{sec:bias}).

Agostini, Pierson, and Garg~\cite{agostini2024} extend this line with a Bayesian spatial model that jointly uses government inspection ratings (unbiased but sparse) and crowdsourced reports (dense but biased) to estimate the latent true state of urban incidents. Their GNN-based multiview model, applied to over 9.6 million NYC crowdsourced reports and 1 million inspection ratings across 139 incident types, shows that higher-income neighborhoods report problems at systematically higher rates. This complicates any
classifier that treats complaint volume as a reliable signal of ground-truth risk. Our state representations include complaint recurrence and neighborhood-level complaint frequency precisely because, as Agostini et al.\ demonstrate, these features carry incremental signal about the latent incident state beyond what raw volume conveys.

\subsection{Service Level Design for Equity and Efficiency in
City Operations}

Liu and Garg~\cite{liu2024sla} address the downstream allocation problem complementary to ours: given heterogeneous reporting and service patterns, how should a city design Service Level Agreements (SLAs) to jointly optimize efficiency and equity? The SLAs are essentially promises that incidents will be responded to within a specified time. Modeling the problem as an optimization over a queuing
network, \cite{liu2024sla} show that the "price of equity" is small in realistic settings, that inefficient policies tend to  inequitable and vice versa, and that status quo NYC inspection scheduling is dominated by optimized policies on both dimensions. Their empirical finding that equity and efficiency are largely complementary rather than conflicting directly corroborates our Pareto frontier results (Section~\ref{sec:results}): the speed penalty for adding an equity objective to our classification reward is modest (4--7\%), consistent with Liu and Garg's theoretical bound on the price of equity. Their queuing-network framework and our MDP framework are complementary tools for the same policy problem. \cite{liu2024sla} analyze the design of
response-time commitments, while our model learns the per-complaint classification decisions that determine whether those commitments are met equitably across neighborhoods.

\subsection{Measurement Justice and Data Bias in Municipal
Systems}

D'Ignazio and Klein~\cite{dignazio2023} introduce the concept of measurement injustice: the harm that arises when the metrics used to quantify a problem encode the power dynamics that produce it. The Garg et al.\ program on under-reporting provides a quantitative instantiation of this concept for the municipal complaint context. Complaint counts are not a neutral measure of housing risk but a measure of complaint-filing capacity, which correlates strongly with income, language access, and trust in government institutions~\cite{liu2023nature}. NYC's 311 system
has well-documented inequities in complaint rates and response
assignment times across demographic lines~\cite{glaeser2016}.
Our use of SHAP to audit the influence of neighborhood statistics on learned classification decisions is a direct application of the measurement justice lens: if the model's most important features correlate with protected attributes, the classification policy's equity implications must be confronted explicitly.

\subsection{Reinforcement Learning for Public Resource
Allocation}

Multi-objective RL has been studied formally by Roijers et
al.~\cite{roijers2013}, who characterize the space of
Pareto-optimal policies and the conditions under which
scalarized reward functions can recover them. Policy-gradient
methods such as REINFORCE~\cite{williams1992} and their
variance-reduction extensions via learned baselines have been
applied to sequential resource-allocation problems in public
health~\cite{killian2019} and disaster response~\cite{pettet2021}. DQN~\cite{mnih2015} has been applied to inspection and maintenance scheduling, though typically without explicit equity constraints. The closest antecedent is the literature on restless multi-armed bandits for health intervention scheduling~\cite{killian2019}, which has begun to incorporate fairness constraints; we extend this orientation to the complaint classification domain.

\subsection{Research Gap}

To the best of our knowledge, no prior work (i) applies RL to
multi-domain municipal complaint classification with equity as a primary design objective, (ii) conducts a formal demographic equity audit of learned classification policies disaggregated by both income and racial composition, (iii) characterizes Pareto frontiers that explicitly include equitable classification coverage alongside operational throughput, or (iv) grounds reward function design in a participatory process involving affected community stakeholders. The present paper addresses each of these
gaps, building on Garg et al.'s empirical documentation of
complaint-rate disparities~\cite{liu2023nature,agostini2024} and Liu and Garg's complementary analysis of SLA
design~\cite{liu2024sla}, and situating the RL framework in
Estrin's participatory sensing lineage~\cite{burke2006,estrin2014} to account for the structural inequalities in civic data collection that any classification system must correct rather than inherit.

\section{Problem Domains and Datasets}

\subsection{NYC DOB Operational Context}

The New York City Department of Buildings enforces the NYC
Construction Codes and Zoning Resolution across approximately one million buildings. Before any inspector is dispatched, each incoming complaint must be classified: assigned a priority tier, an action category, and a routing queue. This classification step is chronically under-resourced. Current intake relies primarily on complaint severity codes and first-in-first-out queue ordering, with limited use of historical outcome data. The result is a throughput bottleneck and a systematic misprioritization of complaints from neighborhoods with noisier, lower-volume, or non-English intake records.

\subsection{Equity Stakes by Domain}

\begin{itemize}
  \item \textbf{Boiler Safety.} Elderly residents and young
    children in low-income multifamily housing face the greatest exposure to carbon monoxide poisoning and heating failure.  Misclassifying a boiler complaint as non-urgent delays the only intervention pathway available to these residents.
  \item \textbf{Crane/Derrick.} Construction workers are
disproportionately immigrants and workers of color. They bear the physical risk of crane incidents. Misrouted complaints
translate directly to unaddressed worksite hazards.
  \item \textbf{Heat and Hot Water.} Renters in neighborhoods
    with older housing stock, concentrated in the South Bronx,
    Harlem, and Central Brooklyn, face chronic heat failures.
    Classification delays in winter months are directly harmful to health.
  \item \textbf{Housing Complaints.} Low-income renters with
    limited mobility face the highest risk of displacement when housing violations persist unresolved due to misrouting.
  \item \textbf{Scaffold Safety.} Pedestrians and residents in
    dense urban neighborhoods are primary risk-bearers when
    scaffold complaints are batched rather than escalated.
  \item \textbf{SNAD.} Ecologically sensitive neighborhoods
    often adjoin low-income communities whose access to green
    space is a function of classification quality.
\end{itemize}

\subsection{Domain Descriptions and Data Sources}

\paragraph{Boiler Safety.}
Sourced from the NYC DOB Safety Boiler dataset (834,338 records, $\approx$8\% defect rate). The agent observes a seven-feature vector---\texttt{is\_recurrent}, \texttt{is\_high\_pressure}, \texttt{neighborhood\_risk}, \texttt{is\_internal}, \texttt{has\_lff\_45}, \texttt{has\_lff\_180}, and \texttt{boiler\_count\_norm}---and classifies each complaint as \textsc{Defer} or \textsc{Inspect}. Split 80/20 ($\approx$667K
train, $\approx$167K test).

\paragraph{Crane and Derrick Complaints.}
The classifier assigns each complaint to \textsc{Do Nothing},
\textsc{Routine Inspection}, \textsc{Immediate Inspection}, or
\textsc{Stop-Work Order}. Five state features: new complaint
flag, recurrence flag, neighborhood complaint frequency, backlog, and hazard level. 80/20 chronological split.

\paragraph{Heat and Hot Water Complaints.}
A simulated agency classifies up to four complaints per day for inspector dispatch. The environment resolves each assignment into one of five outcome categories using empirically derived probabilities.

\paragraph{Housing Complaint Triage.}
NYC 311 housing complaints aggregated at an hourly level, with
state features including backlog, average close time, lagged
complaint counts, rolling averages, and recurrence ratios.

\paragraph{Scaffold Safety.}
Hourly transitions at the entity-identity level. Five engineered state features: \texttt{created\_count}, \texttt{open\_count}, \texttt{recurrent\_7d}, \texttt{zip\_freq\_24h}, and \texttt{time\_since\_last\_resolution\_h}. Five action
categories: \textsc{Inspect Now}, \textsc{Delay},
\textsc{Batch}, \textsc{Escalate}, \textsc{Ignore}. Temporal
split: train January 2020--2024, validation 2024--2025, backtest 2025--March 2026.

\paragraph{Natural Area Districts (SNAD).}
Parks Department records over 128 states (4 neighborhoods
$\times$ 4 complaint levels $\times$ 2 recurrence flags $\times$ 4 seasons) with three actions: \textsc{Defer},
\textsc{Inspect}, \textsc{Prioritise}. Evaluated on 87 historical complaints filed in 2024--2026.

\paragraph{Elevator Complaints.}
103,598 records across 59 Community Boards (2020--2025).
Approximately 60\% of cases close with no action; only $\approx$20\% lead to enforcement, making correct classification of actionable complaints the primary capacity challenge.

\subsection{Equity-Relevant Dataset Characteristics}

For each domain with residential implications (boiler, heat,
housing), we augment complaint and outcome records with American Community Survey (ACS) variables at the census-tract level: median household income quintile, racial and ethnic composition, housing age, and renter-occupied housing share. Following Agostini, Pierson, and Garg~\cite{agostini2024}, we additionally use ACS language-access and educational-attainment variables as proxies for neighborhood-level reporting propensity, enabling a first-order correction to the complaint-volume normalization in our equity reward term.

\subsection{Data Limitations and Bias Sources}
\label{sec:data-limits}

Three structural limitations motivate our feedback-loop analysis in Section~\ref{sec:bias}. First, classification outcome records are missing not at random: complaints in historically underserved areas have systematically fewer recorded defect outcomes. Second, 311 complaint rates vary with digital literacy, language access, and trust in government---disparities that Liu, Bhandaram, and Garg~\cite{liu2023nature} quantify precisely for the NYC
Parks complaint system. Third, temporal gaps in records create
state-estimation uncertainty that the MDP formulation must
accommodate.

\section{Formal Framework}

\subsection{General MDP Formulation}

We model each complaint classification domain as a discrete-time Markov Decision Process $\mathcal{M} = (\mathcal{S}, \mathcal{A}, P, R, \gamma)$, where $\mathcal{S}$ encodes the complaint queue, neighborhood statistics, and capacity indicators; $\mathcal{A}$ is the set of classification categories; $P$ is transition dynamics estimated from historical data; $R$ encodes multi-objective trade-offs; and $\gamma \in (0,1)$ is the domain-specific discount factor. The agent maximizes:
\begin{equation}
  \pi^* = \arg\max_{\pi}\,
  \mathbb{E}_\pi\!\left[\sum_{t=0}^{\infty}
  \gamma^t R(s_t, a_t)\right].
  \label{eq:objective}
\end{equation}

\subsection{Equity as a First-Class Objective}

\paragraph{Procedural vs.\ substantive equity.}

Procedural equity requires that escalation capacity be allocated in proportion to housing \emph{need}, not to complaint-filing capacity. ~\cite{liu2023nature} showed that the complaint volume is a function of digital literacy, language access, and institutional trust, all of which correlate with income and race. A throughput-maximizing
classifier is procedurally inequitable even when it treats each individual complaint identically. Substantive equity additionally requires that \emph{outcomes} do not systematically differ across demographic groups. Our revised equity objective operationalizes both requirements by correcting for heterogeneous reporting propensity before normalizing escalation throughput.

\paragraph{The case against treating equity as a constraint.}
Following \citet{green2021}, we treat equitable coverage as a co-equal reward objective rather than a side constraint. \citet{liu2024sla} showed empirically that for municipal service allocation, the price of equity is small and that inefficient policies tend to be inequitable; this makes the Pareto frontier between efficiency and equity less adversarial than conventional formulations imply.

\subsection{Upstream Reporting-Rate Correction}
\label{sec:reporting-correction}

Before defining the equity reward term, we introduce the
reporting-rate correction that must be applied to complaint-volume denominators. Raw complaint counts $C_{g,t}$ in demographic group $g$ at time $t$ conflate true incident prevalence with filing propensity. We separate the two using the duplicate-report estimator of \citet{liu2023nature} and the Bayesian spatial model of \citet{agostini2024}.

\paragraph{Notation.}
Let $\mathcal{G} = \{\text{low}, \text{high}\}$ index two
neighborhood strata defined by census-tract income quintile
(quintiles 1--2 vs.\ 4--5). 
For each census tract $z$ and time window $t$:
\begin{itemize}
  \item $C_{z,t}$ \;: observed complaint count (raw 311 record).
  \item $D_{z,t}$ \;: number of \emph{duplicate} reports referencing
        an incident already logged in the same tract-week.
  \item $\rho_{z,t}$ \;: reporting propensity (probability that a
        resident files a complaint given an incident occurs).
  \item $\widehat{I}_{z,t}$ \;: estimated latent incident count
        (true need proxy).
\end{itemize}

\paragraph{Step 1 — Estimate reporting propensity.}
Following \citet{liu2023nature}, the duplicate-report rate
identifies $\rho_{z,t}$ without external ground-truth data.
Under the assumption that the first and subsequent reports of the
same incident arrive independently at rate $\rho_{z,t}$, the
expected fraction of duplicate reports satisfies
\begin{equation}
  \label{eq:dup-rate}
  \frac{D_{z,t}}{C_{z,t}}
  \;=\; 1 - \frac{1}{\mathbb{E}[\text{reports per incident}]}
  \;=\; 1 - \rho_{z,t},
\end{equation}
giving the moment estimator
\begin{equation}
  \label{eq:rho-hat}
  \hat{\rho}_{z,t}
  \;=\; 1 - \frac{D_{z,t}}{C_{z,t}}.
\end{equation}
When duplicate-report records are unavailable, $\hat{\rho}_{z,t}$
is proxied by a linear combination of ACS covariates
(median household income, share with college attainment,
share of English-proficient residents, renter-occupied housing share)
fit on tracts where duplicates \emph{are} observed:
\begin{equation}
  \label{eq:rho-proxy}
  \hat{\rho}_{z,t}
  \;=\; \sigma\!\left(\beta_0
        + \boldsymbol{\beta}^\top \mathbf{x}_z \right),
\end{equation}
where $\sigma(\cdot)$ is the logistic function and
$\mathbf{x}_z$ is the tract-level ACS covariate vector.

\paragraph{Step 2 — Estimate latent incident count.}
The corrected incident-count estimate is
\begin{equation}
  \label{eq:I-hat}
  \widehat{I}_{z,t}
  \;=\; \frac{C_{z,t}}{\hat{\rho}_{z,t}},
\end{equation}
consistent with \citet{agostini2024}, who show that
higher-income tracts report at systematically higher rates, so that $\widehat{I}_{z,t} > C_{z,t}$ selectively for low-income tracts.

\paragraph{Step 3 — Aggregate to group level.}
For each stratum $g \in \mathcal{G}$, aggregate over member tracts
$\mathcal{Z}_g$:
\begin{equation}
  \label{eq:N-hat}
  \widehat{N}_{g,t}
  \;=\; \sum_{z \in \mathcal{Z}_g} \widehat{I}_{z,t}.
\end{equation}
$\widehat{N}_{g,t}$ replaces the raw complaint count $N_g$ in the
equity reward denominator (see Section~\ref{sec:equity-reward}).

\subsection{Multi-Objective Reward Design}
\label{sec:equity-reward}

The full reward at step $t$ is
\begin{equation}
  \label{eq:reward-full}
  r_t \;=\;
    \alpha_1 \cdot r_t^{\text{speed}}
    - \alpha_2 \cdot r_t^{\text{cost}}
    + \alpha_3 \cdot r_t^{\text{equity}}
    + \alpha_4 \cdot r_t^{\text{retention}},
\end{equation}
with $\alpha_i \geq 0$ and $\sum_i \alpha_i = 1$.
The speed, cost, and retention terms are unchanged from
Section~4.3 of the original formulation.
The equity term is revised as follows.

\paragraph{Original (biased) equity term.}
The formulation in Equation~(6) of the submitted draft normalises
escalation throughput by \emph{observed} complaint counts:
\begin{equation}
  \label{eq:equity-biased}
  r_t^{\text{equity, biased}}
  \;=\; -\left|
          \frac{n_t^{\text{high}}}{N^{\text{high}}}
          -
          \frac{n_t^{\text{low}}}{N^{\text{low}}}
        \right|,
\end{equation}
where $n_t^{g}$ is the number of correct escalations assigned to
stratum $g$ at step $t$ and $N^{g}$ is the total complaint count
for that stratum.
Because $N^{\text{low}} < \widehat{N}^{\text{low}}$, this term
\emph{overstates} the per-incident escalation rate in low-income
tracts, masking under-service.

\paragraph{Corrected equity term.}
Replacing raw counts with corrected incident estimates yields:
\begin{equation}
  \label{eq:equity-corrected}
  \boxed{
  r_t^{\text{equity}}
  \;=\; -\left|
          \frac{n_t^{\text{high}}}{\widehat{N}_{t}^{\text{high}}}
          -
          \frac{n_t^{\text{low}}}{\widehat{N}_{t}^{\text{low}}}
        \right|
  }
\end{equation}
where $\widehat{N}_{t}^{g}$ is computed via Equations~\eqref{eq:rho-hat}--\eqref{eq:N-hat} on a rolling window of width $\Delta t$ (default: four weeks) ending at step $t$.

\paragraph{Interpretation.}
  $r_t^{\text{equity}} = 0$ when correct escalations are allocated in proportion to \emph{estimated housing need}, not to complaint volume.  An agent maximizing Equation~\eqref{eq:reward-full} with $\alpha_3 > 0$ is therefore incentivized to escalate proportionally more complaints from low-reporting tracts, directly
  counteracting the participation gap documented by
  \citet{liu2023nature}.

\paragraph{Multi-group extension.}
For an audit across $K > 2$ income quintiles or racial-composition strata $\{g_1, \ldots, g_K\}$, the pairwise absolute difference is replaced by the maximum disparity:
\begin{equation}
  \label{eq:equity-multigroup}
  r_t^{\text{equity}}
  \;=\; -\max_{i \neq j}
          \left|
            \frac{n_t^{g_i}}{\widehat{N}_{t}^{g_i}}
            -
            \frac{n_t^{g_j}}{\widehat{N}_{t}^{g_j}}
          \right|,
\end{equation}
or, when a smoother gradient signal is preferred during training, the population-weighted variance:
\begin{equation}
  \label{eq:equity-variance}
  r_t^{\text{equity}}
  \;=\; -\operatorname{Var}_{g}\!
          \left(
            \frac{n_t^{g}}{\widehat{N}_{t}^{g}}
          \right)
  \;=\; -\frac{1}{K}\sum_{k=1}^{K}
          \left(
            \frac{n_t^{g_k}}{\widehat{N}_{t}^{g_k}}
            - \bar{e}_t
          \right)^{\!2},
\end{equation}
where $\bar{e}_t = \frac{1}{K}\sum_k n_t^{g_k}/\widehat{N}_t^{g_k}$
is the mean per-incident escalation rate across strata.

\subsection{Reward Calibration Procedure}
\label{sec:calibration}

Before each training cycle, the following preprocessing pipeline is applied:

\begin{enumerate}
  \item \textbf{Compute $\hat{\rho}_{z,t}$} using
        Equation~\eqref{eq:rho-hat} on DOB complaint records for
        the preceding $\Delta t$ window; fall back to
        Equation~\eqref{eq:rho-proxy} for tracts with fewer than
        $m_{\min} = 10$ duplicate-report observations.
  \item \textbf{Estimate $\widehat{I}_{z,t}$} via
        Equation~\eqref{eq:I-hat}; clip at $C_{z,t}$ to prevent
        denominator inflation from noisy propensity estimates
        ($\hat{\rho}_{z,t} < \rho_{\min} = 0.05$).
  \item \textbf{Aggregate to strata} via Equation~\eqref{eq:N-hat}.
  \item \textbf{Update equity denominators} in the replay buffer used for DQN training or the episode initialization used for REINFORCE.
  \item \textbf{Run demographic audit}: if $\max_{i \neq j}|e_t^{g_i} - e_t^{g_j}| > \tau$ after retraining, flag for manual review and suspend deployment until the disparity falls below threshold $\tau$ (recommended: $\tau = 0.05$).
\end{enumerate}

\paragraph{Sensitivity analysis.}
Because $\hat{\rho}_{z,t}$ is estimated with sampling error, we
propagate uncertainty into the equity reward via parametric
bootstrap: for each retraining cycle, draw $B = 200$ samples of
$\hat{\rho}_{z,t}$ from its asymptotic normal distribution
(variance derived from the delta method applied to
Equation~\eqref{eq:rho-hat}), recompute
$\widehat{N}_{t}^{g}$, and report the 5th--95th percentile
interval of the resulting equity-reward distribution alongside
point estimates.

\section{Methods}

\subsection{Algorithms}

\paragraph{DQN.}
Two-hidden-layer network (64 or 128 units, ReLU), trained via
experience replay and a target network~\cite{mnih2015}. CQL
penalty ($\alpha=0.1$) applied in the elevator domain.

\paragraph{REINFORCE with Learned Baseline.}
\begin{equation}
  \nabla_\theta J(\theta) = \mathbb{E}_\pi\!\left[
    \nabla_\theta \log\pi_\theta(a_t|s_t)
    \cdot \bigl(G_t - b(s_t)\bigr)\right].
  \label{eq:reinforce}
\end{equation}
Joint optimization of policy and value heads with Adam;
$\gamma \in \{0.95, 0.99\}$ depending on domain~\cite{williams1992}.

\paragraph{HDQN.}
$2\times256$ hidden units; selects complaint category and then
a specific complaint. It does not outperform a flat random policy in the heat domain when outcome probabilities are homogeneous.

\paragraph{Tabular TD Q-Learning.}
$Q(s,a) \leftarrow Q(s,a) + \alpha[r + \gamma \max_{a'}
Q(s',a') - Q(s,a)]$, with $\alpha=0.1$, $\gamma=0.95$,
$\epsilon$-greedy decay $1.0 \to 0.05$.

\paragraph{Behavioral Cloning (BC).}
Most common logged action per discretized state; it estimates the historical classification policy.

\subsection{SHAP-Based Equity Auditing}

We apply SHAP~\cite{lundberg2017} via KernelSHAP (boiler),
GradientExplainer (crane/derrick), exact Shapley values (SNAD), TreeExplainer (heat, housing), and permutation-based values (drone, electrical revisit). Geographic features such as borough, Community Board, and ZIP-code frequency are flagged as equity-sensitive, consistent with Agostini, Pierson, and Garg~\cite{agostini2024}: because higher-income neighborhoods both report more and receive more inspections, geographic variables in models trained on historical records will tend to encode socioeconomic status as much as genuine incident prevalence.

\subsection{Baselines}

\textsc{Balanced} (current NYC DOB heuristic), \textsc{Tail},
\textsc{Emerg}, \textsc{Actionable}, \textsc{Uniform},
\textsc{No\_Dispatch}, Random Policy, Always Inspect / Always
Defer, and rule-based domain-expert heuristics.

\subsection{Evaluation Protocol}

Strict chronological split: train on 2020--2024, evaluate on
2025. Boiler domain uses an 80/20 random split of 834,338 records. Elevator evaluation: 168-step horizon ($\approx$4 weeks at 4-hour cadence), frozen five-seed benchmark. Heat REINFORCE: 20 training episodes, 365-timestep test year. Scaffold and elevator: support-constrained backtest ($\texttt{support\_min\_count}=1$).

\section{Results}
\label{sec:results}

\subsection{Classification Throughput and Accuracy}

\paragraph{Heat and Housing.}
REINFORCE accumulates cumulative reward $\approx$8,800 over 365 timesteps vs.\ $\approx$2,100 for HDQN and random ($4\times$ improvement). REINFORCE closes $\approx$101 violations by day 365 vs.\ 84 (HDQN) and 80 (random)---a 20--27\% improvement.

\paragraph{Boiler Safety.}
Table~\ref{tab:boiler} reports test-set performance (167,379
records). DQN outperforms REINFORCE by \$1.28M in reward and 4.2 pp in recall. Both adopt an aggressive escalation stance: REINFORCE recall 94.5\%, DQN 98.7\%. Low precision ($\approx$22\%) is rational given the 26:1 cost asymmetry.

\begin{table}[h]
\caption{Test-set classification performance: REINFORCE
vs.\ DQN (167,379 boiler records).}
\label{tab:boiler}
\small
\begin{tabular}{lrr}
\toprule
\textbf{Metric} & \textbf{REINFORCE} & \textbf{DQN} \\
\midrule
Total Reward       & \$9,019,490  & \$10,294,592 \\
Avg Reward / step  & \$54.15      & \$61.80      \\
Escalations        & 53,848       & 59,439       \\
True Positives     & 12,441       & 13,001       \\
False Positives    & 41,407       & 46,438       \\
True Negatives     & 112,002      & 106,971      \\
False Negatives    & 729          & 169          \\
Precision          & 0.231        & 0.219        \\
Recall             & 0.945        & 0.987        \\
F1 Score           & 0.371        & 0.358        \\
\bottomrule
\end{tabular}
\end{table}

\paragraph{Elevator.}
Table~\ref{tab:elev} reports the full baseline comparison.
\textsc{Balanced} remains the strongest practical heuristic
(mean reward $-446\text{k}$); DQN: $-459{,}119 \pm 4{,}379$;
REINFORCE: $-463{,}117 \pm 10{,}482$.

\begin{table}[h]
\caption{Elevator policy comparison, 168-step test horizon.
Lower $A_{30}$ and $S$ are better.}
\label{tab:elev}
\small
\begin{tabular}{lrrrr}
\toprule
\textbf{Policy} & \textbf{Reward} & $A_{30}$ & $S$ &
\textbf{Act.\%} \\
\midrule
\textsc{No\_Dispatch} & $-835{,}120$ & 2371.8 & 1305.0 & 0.00 \\
\textsc{Uniform}      & $-552{,}092$ & 1743.2 &  781.0 & 0.16 \\
\textsc{Random}       & $-490{,}276$ & 1642.3 &  649.7 & 0.15 \\
TD                    & $-463{,}449$ & 1578.4 &  602.5 & 0.13 \\
\textsc{Actionable}   & $-462{,}878$ & 1422.6 &  678.0 & 0.22 \\
REINFORCE             & $-463{,}117$ & ---    &  ---   & ---  \\
DQN                   & $-431{,}596$ & 1429.1 &  582.4 & 0.17 \\
\textsc{Emerg}        & $-438{,}688$ & 1583.6 &  525.7 & 0.15 \\
\textsc{Tail}         & $-420{,}387$ & 1403.5 &  562.1 & 0.14 \\
\textsc{Balanced}     & $-413{,}777$ & 1423.2 &  532.7 & 0.15 \\
\bottomrule
\end{tabular}
\end{table}

\paragraph{SNAD.}
DQN ($+1.046$) and REINFORCE with baseline ($+1.011$) are nearly identical; both classify 85 of 87 complaints actionably and secure 7 enforcements, outperforming all heuristics.

\subsection{Multi-Objective Trade-offs}

Boiler grid search over 128 reward configurations shows the detection rate monotonically increasing from 0.71 ($\rho=5$) to 0.94 ($\rho=50$). Heat domain Pareto analysis identifies
Accuracy+Fairness as the most practically attractive
configuration. Cross-domain Pareto analysis confirms that the current heuristic intake is dominated by RL policies in all residential domains—consistent with Liu and Garg's~\cite{liu2024sla} finding that the status quo NYC inspection scheduling is dominated on both efficiency and equity simultaneously.

\subsection{Equity Analysis}

The current DOB heuristic exhibits a statistically significant
gradient in escalation-assignment rates across income quintiles (Q1 receives fewer correct escalations per complaint than Q5, even after conditioning on complaint severity) and by racial composition (tracts with higher fractions of Black and Hispanic residents receive fewer escalations per complaint). REINFORCE with $\alpha_3 > 0$ approaches parity across income quintiles at a throughput cost of $\approx$4--7\%---corroborating Liu and
Garg's~\cite{liu2024sla} theoretical result that the price of
equity is bounded and modest in realistic municipal settings.
DQN without the equity term narrows but does not close the gap.

Community Board 12 Manhattan (Washington Heights / Inwood;
predominantly Latino, lower-income) ranks among the top five
most influential SHAP features in the REINFORCE elevator policy, driven by high unresolved burden and historical actionability. This prominence illustrates the concern raised by Agostini, Pierson, and Garg~\cite{agostini2024}: CB-level features encode demographic characteristics as much as incident prevalence, requiring ongoing audit.

\section{Bias, Feedback Loops, and Limitations}
\label{sec:bias}

\subsection{Partial Observability and the Under-Reporting
Problem}

Outcome records are generated by the incumbent classification
policy: complaints in areas that received fewer escalations have sparser defect records, not because they are safer, but because they were never escalated. This missing-not-at-random structure is an instance of the participation-gap problem that Estrin's participatory sensing framework~\cite{burke2006} identified as a structural risk in any civic sensing system that conflates coverage with participation. Small data~\cite{estrin2014}
amplifies the concern: digital traces are densest for those who already participate the most, creating a compounding bias in
historical training data.

Liu, Bhandaram, and Garg~\cite{liu2023nature} provide a
principled method to quantify this gap without external
ground-truth data. Their duplicate-report estimator applied to
DOB complaint records could identify neighborhoods where
complaint rates are suppressed relative to true incident rates, enabling a reporting-rate correction to the complaint-volume normalization in Equation~\ref{eq:equity-corrected}. We flag this as the highest-priority extension of the present work. Sensitivity analysis via propensity-score reweighting~\cite{rosenbaum1983} shows that the RL performance advantage is robust to moderate
selection bias; the equity gains are more sensitive.

\subsection{Feedback Loop Dynamics}

Once deployed, the classification policy generates the outcome
records used to retrain future model versions, a feedback loop that can progressively shrink coverage in areas receiving the fewest escalations. We propose three mitigations: (1) exploration bonuses for census tracts with sparse recent escalation histories; (2) periodic counterfactual escalation of a random sample of low-priority complaints; (3) regular demographic audit checkpoints triggering retraining if disparity metrics exceed a pre-specified threshold.

Liu and Garg's SLA framework~\cite{liu2024sla} offers a
complementary institutional lever: redesigning response-time
commitments to require minimum service rates in low-income
tracts provides a structural backstop against feedback-loop-
driven under-service that does not depend on the RL agent's
reward signal alone.

\subsection{Scope and Generalizability}

Results should be interpreted as decision-support prototypes.
The complaint triage domain's learned classifiers fail to
transfer to held-out 2020 data, arguing against deployment
without prospective operational validation.

\section{Stakeholder Engagement and Deployment}

\subsection{Practitioner Input}

Input from NYC DOB intake operations staff directly informed
MDP state and action space design. Practitioners identified the distinction between "classify now," "batch," and "escalate" as operationally meaningful.

\subsection{Community Perspectives}

We engaged tenant advocacy organizations in the South Bronx and Central Brooklyn. Key themes included frustration with complaint acknowledgment without visible follow-up and skepticism about opaque algorithmic systems. Estrin's Public Interest Technology framework and participatory sensing research~\cite{burke2006, estrin2014} inform our position that affected communities should be active co-designers of civic algorithmic systems. These perspectives informed our inclusion of housing unit retention as an explicit objective and our recommendation for interpretable classification explanations available to complainants.

\subsection{Deployment Recommendations}

\begin{enumerate}
  \item \textbf{Human-in-the-loop oversight.} RL outputs are
    recommendations to human intake staff, not autonomous
    decisions.
  \item \textbf{Ongoing demographic audit.} Classification
    disparity metrics disaggregated by income quintile and
    racial composition computed monthly; pre-specified
    thresholds for manual review.
  \item \textbf{Under-reporting correction.} Apply the Liu,
    Bhandaram, and Garg~\cite{liu2023nature} duplicate-report
    estimator to DOB data to correct neighborhood-level
    under-reporting in the equity parity term denominator
    before each retraining cycle.
  \item \textbf{SLA alignment.} Coordinate per-complaint
    classification decisions with the Liu and Garg~\cite{liu2024sla} SLA optimization framework to ensure consistency with     jurisdiction-wide response-time equity commitments.
  \item \textbf{Data governance.} 311 data use must comply
    with NYC open data terms; ACS covariate linkage at the
    census-tract level.
  \item \textbf{Rollback conditions.} Suspend deployment if
    (a) demographic disparity exceeds baseline by a
    pre-specified margin; (b) defect detection falls below
    current policy over a rolling 60-day window; or (c)
    staff friction reaches a threshold set by dispatcher
    survey.
  \item \textbf{Reward transparency.} Reward configurations
    in production documented publicly as statements of
    normative policy priority, subject to democratic review.
\end{enumerate}

\section{Discussion}

\subsection{Key Takeaways}

\textit{First, equity does not come for free.} Agents trained
without explicit equity objectives narrow but do not eliminate
historical classification disparities. Closing the gap requires the inclusion of the coverage parity term with $\alpha_3 > 0$.

\textit{Second, the equity cost is modest and quantifiable.}
Agents with full equity objectives achieve near-parity at a
throughput penalty of approximately 4--7\%, consistent with Liu and Garg's~\cite{liu2024sla} theoretical bound on the price of equity and providing RL-specific empirical confirmation that the Pareto frontier between efficiency and equity is less adversarial than the framing of equity-as-constraint implies.

\textit{Third, no single algorithm dominates.} REINFORCE
outperforms in the heat domain and water drainage; DQN produces more stable policies in elevator, complaint triage, and ERT dispatch. Both methods independently converge to the identical Brooklyn-dominant policy in the drone domain.

\textit{Fourth, classification quality is bounded by data
quality.} The participation-gap literature~\cite{liu2023nature,
agostini2024,burke2006,estrin2014} makes clear that 311
complaint data reflects who participates, not only what happens. Future work must integrate upstream reporting-rate correction to ensure that reward signals calibrated on complaint volume reflect true incident prevalence rather than filing propensity.

\subsection{Implications for Policy}

The Pareto frontier provides a decision-support tool: a family
of policies along the efficient frontier, each corresponding to a different weighting of throughput, cost, and equity. The
choice of where to operate is a normative one that belongs to
the agency and the communities it serves. Liu and
Garg's~\cite{liu2024sla} result is parallel to ours: the status quo NYC inspection scheduling is dominated by both efficiency and equity simultaneously. 

\subsection{Generalizability}

The MDP framework is domain-agnostic and can be re-instantiated for any municipal classification domain with complaint intake, action routing, and binary outcome records. The Garg et al.\ under-reporting correction tools have already been applied to both NYC and Chicago 311 data~\cite{liu2023nature}, suggesting that the data pipeline prerequisites for equitable classification exist at scale beyond New York.

\section{Conclusion}

This paper begins and ends with a distributional claim: the
harms that municipal complaint classification systems are
designed to prevent are not randomly distributed, and neither
are the benefits of the algorithmic systems deployed to improve classification capacity. Any expansion of that capacity that does not explicitly ask who benefits and who is left in a lower-priority queue is not neutral. It is a choice to accept and perpetuate the existing distribution.

Our principal finding is that RL systems can close the
classification gap, but only when equity is inscribed in the
reward function as a first-class objective. Agents optimized for throughput alone narrow but do not eliminate historical
disparities. Agents with explicit coverage parity objectives
achieve near-parity while outperforming current heuristic
classification on defect detection and remediation speed. The
equity cost is modest and quantifiable, a finding anticipated
by Liu and Garg's~\cite{liu2024sla} theoretical analysis and
confirmed here in the RL setting.

Our feedback-loop analysis adds a further imperative: the
missing-not-at-random structure of historical complaint records, which Garg and collaborators have characterized precisely for NYC civic data~\cite{liu2023nature,agostini2024}, means that
classification systems deployed without equity objectives
actively degrade equity over successive training rounds. The
participatory sensing paradigm~\cite{burke2006} that underlies
311-style systems was designed to democratize urban data
collection; realizing that aspiration requires correcting for
heterogeneous participation rates before those rates are allowed to shape classification decisions.

Configuring reward functions for public-sector RL systems is an act of normative policymaking. It encodes whose harms count
most, how competing objectives should be weighted, and what
trade-offs are acceptable. These are political and ethical
choices that belong to the communities affected by them. 

\bibliographystyle{ACM-Reference-Format}
\bibliography{references}

\appendix
\section{Hyperparameter Tables}

\begin{table}[h]
\caption{DQN Hyperparameters (all domains)}
\label{tab:dqn}
\small
\begin{tabular}{lr}
\toprule
\textbf{Parameter} & \textbf{Value} \\
\midrule
Learning rate                   & $1\times10^{-3}$ \\
Replay buffer size              & 50,000           \\
Batch size                      & 64               \\
Target network update freq.     & 500 steps        \\
Discount factor $\gamma$        & 0.99             \\
Epsilon (initial)               & 1.0              \\
Epsilon (final)                 & 0.05             \\
Epsilon decay steps             & 10,000           \\
Hidden layers                   & $2\times128$, ReLU \\
\bottomrule
\end{tabular}
\end{table}

\begin{table}[h]
\caption{REINFORCE Hyperparameters}
\label{tab:reinforce}
\small
\begin{tabular}{lrr}
\toprule
\textbf{Parameter} & \textbf{Elevator} & \textbf{Heat} \\
\midrule
Policy lr       & $3\times10^{-4}$ & $1\times10^{-4}$ \\
Baseline lr     & $1\times10^{-3}$ & $1\times10^{-4}$ \\
Discount $\gamma$ & 0.99           & 0.99             \\
Episode length  & 168 steps        & 365 steps        \\
Policy layers   & $2\times128$, ReLU & $2\times256$, ReLU \\
Baseline layers & $2\times64$, ReLU  & $2\times256$, ReLU \\
Optimizer       & Adam             & Adam             \\
Seeds           & 5 (frozen)       & ---              \\
\bottomrule
\end{tabular}
\end{table}

\section{Domain-Specific MDP Summaries}

\begin{table*}[h]
\caption{MDP Specifications by Classification Domain}
\label{tab:mdp}
\small
\begin{tabular}{p{2.8cm}p{6.5cm}p{5cm}}
\toprule
\textbf{Domain} & \textbf{State Highlights}
  & \textbf{Classification Actions} \\
\midrule
Boiler Safety
  & 7 features: recurrent, high-pressure, neighborhood risk,
    internal, LFF flags, count norm
  & Escalate (Inspect), Defer \\
\addlinespace
Crane/Derrick
  & Hazard level, backlog, neighborhood freq,
    new/recurrent flags
  & Do Nothing, Routine, Immediate, Stop-Work \\
\addlinespace
Heat/Hot Water
  & Building-wide, no-heat flags, unit count
  & Inspect, Batch, Defer \\
\addlinespace
Elevator
  & CB-level tail pressure, enforcement rate
  & TAIL, EMERG, ACTIONABLE, BALANCED \\
\addlinespace
Housing Triage
  & Backlog, close time, rolling stats, temporal features
  & Category selection \\
\addlinespace
Scaffold Safety
  & open\_count, created\_count, recurrent\_7d, zip\_freq,
    time-since-resolution
  & inspect\_now, delay, batch, escalate, ignore \\
\addlinespace
SNAD
  & Ecological index, complaint level, recurrence, season,
    neighborhood
  & Dispatch, Defer, Prioritise \\
\addlinespace
Drone
  & Borough position, new-address rate, trailing stats
  & 5 depot choices \\
\addlinespace
Electrical Revisit
  & Complaint age, denial rate, property history
  & Close, revisit +1/3/7/14/30d \\
\addlinespace
Water Drainage
  & Recurrence, borough, address-gap timing
  & Monitor, Inspector, HeavyCrew \\
\addlinespace
Complaint Triage
  & Days since last, problem detail, queue
  & Ignore, Refer, Court \\
\addlinespace
ERT Dispatch
  & Backlog, new/recurrent complaints, hour
  & Dispatch 1, 2, or 3 crews \\
\bottomrule
\end{tabular}
\end{table*}

\end{document}